\documentclass[amsmath, amsfonts, superscriptaddress, showpacs, twocolumn, prb]{revtex4}
\usepackage{graphicx}
\usepackage{epsfig}
\usepackage{bm}
\usepackage{dcolumn}
\usepackage{amsmath}
\usepackage{amssymb}
\usepackage{euscript}

\begin{document}

\title{Transport in partially equilibrated inhomogeneous quantum wires}

\author{Alex Levchenko}
\affiliation{Materials Science Division, Argonne National
Laboratory, Argonne, Illinois 60439, USA}

\author{Tobias Micklitz}
\affiliation{Dahlem Center for Complex Quantum Systems and Institut
f\"{u}r Theoretische Physik, Freie Universit\"{a}t Berlin, 14195
Berlin, Germany}

\author{J\'{e}r\^{o}me Rech}
\affiliation{Centre de Physique Th\'{e}orique, UMR 6207, Case 907,
Luminy, 13288 Marseille Cedex 9, France}

\author{K.~A.~Matveev}
\affiliation{Materials Science Division, Argonne National
Laboratory, Argonne, Illinois 60439, USA}

\begin{abstract}
  We study transport properties of weakly interacting
  one-dimensional electron systems including on an equal footing
  thermal equilibration due to three-particle collisions and the
  effects of large-scale inhomogeneities. We show that equilibration
  in an inhomogeneous quantum wire is characterized by the
  competition of interaction processes which reduce the electrons
  total momentum and such which change the number of right- and
  left-moving electrons. We find that the combined effect of
  interactions and inhomogeneities can dramatically increase the
  resistance of the wire.  In addition, we find that the
  interactions strongly affect the thermoelectric properties of
  inhomogeneous wires and calculate their thermal conductance,
  thermopower, and Peltier coefficient.
\end{abstract}

\date{September 8, 2010}


\maketitle

\section{Introduction}

Transport properties of low-dimensional systems have been subject of
intensive research work over the last two decades. One of the
fundamental discoveries that has driven the field was the
observation of conductance quantization in ballistic quantum wires
and quantum point contacts.~\cite{G-QPC-Exp-1,G-QPC-Exp-2} It was
found that conductance exhibits a staircase-like dependence on the
electron density with the universal step. The understanding of this
phenomenon follows already from the single-electron picture, which
predicts for the conductance of a one-dimensional single
channel-clean wire,~\cite{G-QPC-Theor}
\begin{equation}\label{G-0}
G=\frac{2e^2}{h}\,.
\end{equation}
The physical origin of conductance plateaus at certain gate voltages
was associated with a fixed number of occupied electronic subbands,
each supplying one quantum of conductance $2e^2/h$. Within the same
approach of noninteracting particles both charge and energy are
carried by electronic excitations. This results in the universal
relation between electric and thermal conductances, known as the
Wiedemann-Franz law $K=(\pi^2/3e^2)TG$. The thermal conductance of
noninteracting electrons is thus
\begin{equation}\label{K-0}
K=\frac{2\pi^2}{3h}T\,.
\end{equation}
In addition to $G$ and $K$ two thermoelectric coefficients of the
electron gas are usually of great interest. These are thermopower
$S$, which relates an induced voltage drop across the wire to
applied temperature gradient, and Peltier coefficient $\Pi$
connecting electric and heat currents. These two coefficients are
connected by an Onsager relation $\Pi=ST$. In the absence of
interactions the thermopower and Peltier coefficients are
exponentially small
\begin{equation}\label{Pi-S-0}
\Pi=ST\propto e^{-\mu/T}
\end{equation}
at low temperatures $T\ll\mu$. (Here $\mu$ is the chemical
potential.) The reason for such strong suppression of thermopower
and Peltier coefficients is the partial cancellation between heat
currents carried by particles with energies $\mu+\epsilon$ and
$\mu-\epsilon$. Only the absence of electronic states below the
bottom of the band prevents $\Pi$ and $S$ from vanishing exactly.

The remarkable success of the simple single-electron picture in
describing the quantization of conductance and explaining the
temperature dependence of thermoelectric coefficients is attributed
to the fact that quantum wires are always connected to
two-dimensional leads, where interactions between electrons do not
play a significant role. Even though the interactions in the wire
are usually not weak, i.e., $e^2/\hbar v_F\gtrsim1$, where $v_F$ is
the Fermi velocity, it has been shown within the so-called
Luttinger-liquid model of one-dimensional electrons that the
interactions inside the wire do not affect conductance
quantization.~\cite{Maslov-Stone,Ponomarenko,Safi-Schulz} It is no
surprise then that a number of recent
experiments,~\cite{Exp-1,Exp-2,Exp-3,Exp-4,Exp-5,Exp-6,Exp-7,Exp-8,Exp-9}
that revealed deviations from the perfect quantization, \eqref{G-0},
in low-density wires, attracted a great deal of theoretical
attention.~\cite{Theor-1,Theor-2,Theor-3,Theor-4,Theor-5,Theor-6,Theor-7}
These deviations often take the form of a shoulder-like feature,
which develops at finite temperature just below the first quantized
plateau, around $0.7\times2e^2/h$. At present there is no consensus
on the theoretical interpretation of this phenomenon. However, it is
generally accepted that electron-electron interaction effects should
be involved in explaining these experimental observations.

In a number of recent
publications~\cite{Theor-3,Theor-4,Theor-5,Theor-8,Theor-9,Theor-10,
Theor-11,Theor-12,Theor-13,Sirenko,Khodas,Lunde,Jerome-1,Jerome-2,Jerome-3,Tobias,Gutman,Buttiker}
transport properties of one-dimensional conductors were reconsidered
focusing on the physics which lies beyond an ideal Luttinger-liquid
model. In particular, when studying the temperature dependence of
the corresponding kinetic coefficients
Refs.~\onlinecite{Lunde,Jerome-1,Jerome-2,Jerome-3,Tobias,Gutman,Buttiker,Altimiras,Degiovanni}
emphasized one fundamental aspect of interactions, namely the role
of physical processes that lead to equilibration of electrons inside
the wire. It should be emphasized that equilibration is absent in an
ideal Luttinger liquid since bosonic elementary excitations of the
latter have infinite lifetime, thus there is no relaxation towards
equilibrium in these systems, no matter how strong the interactions
are. In higher-dimensional systems equilibration at low temperatures
is primarily provided by pair collisions of electrons. These,
however, do not provide relaxation in one-dimensional systems. This
is due to the conservation laws for momentum and energy which
severely restrict the phase space available for scattering. As a
result, pair collisions in ideal one-dimensional wires can occur
with a zero momentum change or an interchange of the two momenta,
leaving the distribution function unaffected. The leading
equilibration mechanism thus involves collisions of more than two
particles. For a weakly interacting system, it is then natural to
assume that equilibration is provided by three-particle scattering
processes.~\cite{Sirenko,Khodas,Lunde} This, of course, also relies
on the additional assumption that other degrees of freedom, which
can absorb energy and momentum from electrons (phonons, for example)
can be ignored. This assumption is acceptable in many cases since
electron-phonon coupling constant is typically much smaller than
that due to the electron-electron interactions.

In practice, long one-dimensional structures are strongly prone to
inhomogeneities inevitably present due to the nearby gates or
charged dopants underlying the wire. However, most preceding works
studied effect of equilibration on transport assuming uniform
(clean) wires. The notable exceptions include
Refs.~\onlinecite{Jerome-1} and \onlinecite{Jerome-2} where smooth
inhomogeneities were accounted for while assuming full equilibration
of the electronic system. The purpose of the present work is to
study effects of inhomogeneities on transport properties of
partially equilibrated quantum wires. We focus our attention to the
situation where the scale of inhomogeneities $b$ is much larger than
the electron Fermi wavelength, $b\gg\lambda_F$, such that
backscattering of electrons from the inhomogeneities is negligible.
In this case only interactions (three-particle collisions) may
interrupt direct flow of the electron liquid and convert
(backscatter) some right-moving electrons into left-moving ones.

Since non-uniform systems are no longer translationally invariant,
there is an additional scattering mechanism, which can relax
electron momentum without changing the number of right- and
left-moving particles.~\cite{Jerome-1,Jerome-2} We find that
equilibration due to three-particle collisions and
inhomogeneity-induced momentum-nonconserving scattering compete with
each other. An interplay between these two effects leads to a very
interesting picture of the electronic transport and results in
temperature-dependent corrections to the wire resistance and
thermoelectric coefficients.

The paper is organized as follows. In Sec.~\ref{Sec-UnifromWire} we
discuss the general structure of the electron distribution function
and transport in a clean one-dimensional wire. We briefly mention
the resulting transport coefficients for this case, which were
recently reported in Ref.~\onlinecite{Tobias}. In
Sec.~\ref{Sec-Formalism} we develop a general formalism which
enables us to treat equilibration of the electron system due to
three-particle collisions and inhomogeneity-induced scattering on
equal footing. The central part of our work is
Sec.~\ref{Sec-Transport} where we apply this theory to the
calculation of transport coefficients in one-dimensional
inhomogeneous wires. From our general expressions we recover the
results known for the short and long uniform wires, and also
consider several experimentally relevant simple models of
inhomogeneous case. We summarize our results in
Sec.~\ref{Sec-Discussions}. Supplementary appendices accompany some
technical aspects of our calculations.

%------------------------------------------------------------------------------------------------------------
\section{Transport in uniform  quantum wires}\label{Sec-UnifromWire}

In the absence of interactions, left- and right-moving electrons
inside the wire are at equilibrium with the reservoir they
originated from. If a voltage bias $V$ and/or temperature difference
$\Delta T$ is applied between the reservoirs, then corresponding
equilibria differ from each other, giving rise to a particular form
of the nonequilibrium distribution function inside the wire. This
distribution depends on the direction of motion of electrons and for
the right- and left-moving particles is controlled, respectively, by
the left and right lead,
\begin{equation}\label{f-0}
f_p=\frac{\theta(p)}{e^{(\epsilon_p-\mu_l)/T_l}+1}+
\frac{\theta(-p)}{e^{(\epsilon_p-\mu_r)/T_r}+1}\,.
\end{equation}
Here $\epsilon_p=p^2/2m$ is the energy of an electron with momentum
$p$ and $\theta(p)$ is the unit step function. The difference
between the chemical potentials (temperatures) in the leads is equal
to the voltage (temperature difference) applied to the wire
$\mu_l-\mu_r=eV$ ($T_l-T_r=\Delta T$). Using the distribution
function \eqref{f-0} at $\Delta T=0$ one can find electric current
$I=GV$ with the conductance of noninteracting electrons
$G=(2e^2/h)(1+e^{-\mu/T})^{-1}$, thus recovering, Eq.~\eqref{G-0},
up to an exponentially small correction. The same distribution,
Eq.~\eqref{f-0}, provides thermal conductance, Eq.~\eqref{K-0}, and
thermoelectric coefficients $\Pi=TS=(\mu/e)e^{-\mu/T}$, consistent
with Eq.~\eqref{Pi-S-0}.

In the presence of interactions ballistic propagation of electrons
through the wire may be interrupted by collisions with other
electrons. As a result of these collisions, some electrons change
their direction of motion thus losing memory of the lead they
originated from. Such backscattering processes modify the electron
distribution function which is then no longer given by
Eq.~\eqref{f-0}. It is important to realize that the effect of
electron collisions on the distribution function depends strongly on
the length of the wire. Indeed, electrons traverse short wires
relatively fast, such that interactions do not have time to change
distribution, Eq.~\eqref{f-0}, considerably. On the other hand, in
the limit of very long wire one should expect full equilibration of
left- and right-moving electrons into a single distribution, even in
the case of weak interactions.

For the Galilean invariant system one can easily infer the electron
distribution function in a fully equilibrated state. Indeed, viewed
from a reference frame moving with the drift velocity $v_d=I/ne$
(where $I$ is the electric current and $n$ is the electron density)
the electron system is at rest and must be described by the
equilibrium Fermi distribution. Performing a Galilean transformation
back into the stationary frame of reference this distribution takes
the form,
\begin{equation}\label{f-eq}
f_p=\frac{1}{e^{(\epsilon_p-v_dp-\mu_{\mathrm{eq}})/T_{\mathrm{eq}}}+1}\,,
\end{equation}
where the chemical potential $\mu_{\mathrm{eq}}$ and temperature
$T_{\mathrm{eq}}$ inside the equilibrated wire are, in general,
different from $\mu_{l(r)}$ and $T_{l(r)}$.

At zero temperature, $T=T_{\mathrm{eq}}=0$, the distributions,
Eqs.~\eqref{f-0} and \eqref{f-eq}, coincide, provided
$\mu_{l(r)}=\mu_{\mathrm{eq}}\pm v_dp_F$, where $p_F=\pi\hbar n/2$
is the Fermi momentum of the system. At non-zero temperature the
distribution function, Eq.~\eqref{f-eq}, of electrons inside the
equilibrated wire is slightly different from the distribution,
Eq.~\eqref{f-0}, supplied by the leads. The mismatch between the two
distribution functions results in additional resistance, reducing
the conductance of noninteracting electrons to~\cite{Jerome-3}
\begin{equation}\label{G-eq}
G_{\mathrm{eq}}=\frac{2e^2}{h}\left[1-\frac{\pi^2}{12}\frac{T^2}{\mu^2}\right]\,.
\end{equation}
This result is universal since it was obtained without making any
specific assumptions regarding the process of equilibration. The
corresponding derivation relied uniquely on the analysis of
conservation laws for energy, momentum, and particle number. It is
applicable as long as wire length $L$ exceeds certain equilibration
length $\ell_{\mathrm{eq}}$ such that the distribution,
Eq.~\eqref{f-eq}, is already established. The exact definition of
$\ell_{\mathrm{eq}}$ is model specific and depends on the
interaction between electrons. Quite generically, however, it can be
argued that this length is exponentially large at low temperature
$\ell_{\mathrm{eq}}\propto e^{\mu/T}$. The exponential scale can be
understood from the mechanism of equilibration~\cite{Tobias} which
is also discussed later in the text.

The thermal conductance of fully equilibrated wire is zero
\begin{equation}\label{K-eq}
K_{\mathrm{eq}}=0\,.
\end{equation}
This result can be understood simply from the structure of the
distribution function \eqref{f-eq}. One should recall that thermal
conductance is defined under the condition that electric current
$I=env_d$ vanishes. Thus $v_d=0$ and the distribution,
Eq.~\eqref{f-eq}, takes the form of the standard Fermi-Dirac
distribution. Due to its symmetry $p\to-p$ the heat current carried
by electrons vanishes, regardless of the temperature bias $\Delta T$
applied to the wire. One therefore finds that in an infinitely long
wire the thermal conductance is zero.  One should note, however,
that thermal conductivity is finite,\cite{Tobias} i.e., at
$L\to\infty$ the thermal conductance scales as $K\propto 1/L$.

Unlike the electric and thermal conductances, thermopower and
Peltier coefficients are significantly enhanced by equilibration
effects.  Specifically, $\Pi$ grows from the exponentially small
value, Eq.~\eqref{Pi-S-0}, for short wires $L\ll\ell_{\mathrm{eq}}$
to
\begin{equation}\label{Pi-S-eq}
\Pi_{\mathrm{eq}}=TS_{\mathrm{eq}}=\frac{\pi^2}{6e}\frac{T^2}{\mu}\,,
\end{equation}
in fully equilibrated (long) wires, $L\gg\ell_{\mathrm{eq}}$.

A more careful treatment of the equilibration effects is required in
wires of intermediate length $L\sim \ell_{\mathrm{eq}}$, where the
electron distribution function is only partially
equilibrated.~\cite{Tobias} As we already mentioned, in the case of
weakly interacting electrons the leading mechanism of equilibration
is provided by three-particle collisions.  At low temperatures one
should consider two types of such collisions.  The strongest
scattering events involve three particles near the Fermi level (for
example, one left mover that scatters off two right movers such that
all particles preserve their direction of motion). These collisions
are relatively fast, and the corresponding scattering length
$\ell_{\mathrm{t}}$ scales as a power law of temperature. However,
these collisions alone can not establish the distribution
\eqref{f-eq} since they conserve the number of right- and
left-moving particles. The other important three-particle collisions
involve backscattering of, say, a right-moving electron into a
left-moving one. This backscattering occurs near the bottom of the
band and provides equilibration between the chemical potentials of
right and left movers, thus establishing the distribution,
Eq.~\eqref{f-eq}.  Since at low temperatures the probability to find
an empty state at the band bottom is exponentially small, the
corresponding relaxation process is very slow, and equilibration
length is large, $\ell_{\mathrm{eq}}\propto e^{\mu/T}$.

Let us consider now a segment of the wire, whose length $\Delta L$
is small compared to the equilibration length $\ell_{\mathrm{eq}}$
but large as compared to $\ell_{\mathrm{t}}$, namely,
$\ell_{\mathrm{t}}\ll\Delta L\ll\ell_{\mathrm{eq}}$. This condition
implies that typical electron with energy near the Fermi level
passes through the segment without backscattering so that the
distribution, Eq.~\eqref{f-eq}, cannot be established. On the other
hand, $\Delta L$ is already sufficiently large for electrons to
experience other multiple collisions which allow momentum and energy
exchange between right- and left-moving electrons. Under these
conditions, the electron distribution function in the segment
achieves a state of partial equilibration, in which the numbers
$N^L$ and $N^R$ of the right- and left-moving electrons are
conserved independently. The form of this distribution can be
obtained from the general statistical mechanics argument by
maximizing the entropy of electrons while preserving $N^{L(R)}$,
total energy and momentum of the system,~\cite{Tobias}
\begin{equation}\label{f-part-eq}
f_p=\frac{\theta(p)}{e^{(\epsilon_p-up-\mu^R)/\bar{T}}+1}+
\frac{\theta(-p)}{e^{(\epsilon_p-up-\mu^L)/\bar{T}}+1}\,.
\end{equation}
Here $\bar{T}$ is the effective temperature, parameter $u$ has
dimension of velocity and accounts for the conservation of momentum
in electron collisions, and $\mu^{L(R)}$ are the chemical potentials
of the left- and right-moving particles. The distribution,
Eq.~\eqref{f-part-eq}, smoothly interpolates between the regimes of
no equilibration, Eq.~\eqref{f-0}, and that of full equilibration,
Eq.~\eqref{f-eq}. In the absence of temperature difference, $\Delta
T=0$, the unperturbed distribution, Eq.~\eqref{f-0}, is obtained
from Eq.~\eqref{f-part-eq} by setting $u=0$ and identifying the
chemical potentials with those in the leads: $\mu^R=\mu_l$ and
$\mu^L=\mu_r$. [Here and throughout the paper we use $l(r)$ to
denote left (right) lead while $L(R)$ denote left (right) movers.]
The fully equilibrated distribution, Eq.~\eqref{f-eq}, is obtained
from Eq.~\eqref{f-part-eq} by setting $\Delta\mu=\mu^R-\mu^L=0$. In
this case the electric current is expressed as $I=enu$, which
identifies parameter $u$ as the drift velocity $v_d$. We should
emphasize here that since the distribution, Eq.~\eqref{f-part-eq},
is applicable for the segment of the wire outlined above then all
four parameters $\bar{T}(x), u(x),$ and $\mu^{L/R}(x)$ defining
$f_p$ are, in principle, coordinate dependent.

The implications of the distribution, Eq.~\eqref{f-part-eq}, for the
transport coefficients of partially equilibrated clean wires were
discussed in Ref.~\onlinecite{Tobias}. In the following we
generalize the above picture of electronic transport in
one-dimensional wires accounting for possible non-uniformities of
the system.

\begin{figure}
  \includegraphics[width=8cm]{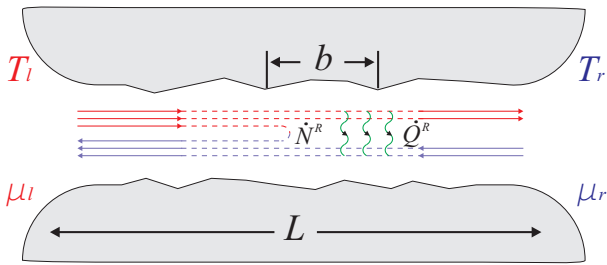}
  \caption{[Color online] Inhomogeneous quantum wire connected
    adiabatically to two-dimensional leads and biased by the small
    voltage $\mu_l-\mu_r=eV$ and/or temperature difference
    $T_l-T_r=\Delta T$.  We assume that the spatial scale $b$ of the
    inhomogeneities is large compared to the Fermi wavelength, and
    thus electrons do not experience backscattering from
    inhomogeneities.  In this case only three-particle equilibration
    processes may interrupt the flow of right-moving electrons and
    convert some right movers into left movers. Wavy lines represent
    heat exchange $\dot{Q}^R$ between right movers and
    left movers.}\label{Fig-Geometry}
\end{figure}

%------------------------------------------------------------------------------------------------------------
\section{Transport in inhomogeneous  quantum wires}\label{Sec-Formalism}

\subsection{Boltzmann equation}\label{Sec-Formalism-BE}

Consider an inhomogeneous quantum wire of length $L$, connected by
ideal reflectionless contacts to noninteracting leads and biased by
a small voltage $V$ and/or temperature difference $\Delta T$, see
Fig.~\ref{Fig-Geometry}. If the spatial variations related to
inhomogeneities occur on a length scale $b$ much larger than the
Fermi wavelength $\lambda_F$, electrons do not suffer any
backscattering. Since the physical picture of equilibration in
one-dimensional wire can be readily understood at the level of
weakly interacting electrons we restrict our attention to this case
and describe the system in the framework of kinetic equation. In
this case electron distribution function $f(t,x,p)$ obeys the
Boltzmann equation,
\begin{equation}\label{KE}
\partial_{t}f+v_p\partial_{x}f-\partial_{x}U(x)\partial_{p}f
=\mathcal{I}\{f\}\,,
\end{equation}
where static potential $U(x)$ accounts for inhomogeneities of the
wire and $\mathcal{I}\{f\}$ conventionally stands for the collision
integral. We are interested in the steady-state regime when
distribution function $f(t,x,p)$ does not depend explicitly on time,
and thus set $\partial_t f=0$. It will be also convenient to split
the distribution $f(x,p)$ into two parts corresponding to the right
and left movers,
\begin{equation}\label{f=fR+fL}
f(x,p)=\theta(p)f^R(x,\epsilon_p(x))+\theta(-p)f^L(x,\epsilon_p(x))\,,
\end{equation}
and express it as the function of energy $\epsilon_p(x)=p^2/2m+U(x)$
for the given momentum $p$. Kinetic Eq.~\eqref{KE} should be
supplemented by the boundary conditions at the ends of the wire that
are controlled by the leads,
\begin{subequations}\label{f-RL-boundary}
\begin{equation}
f^{R}(l,\epsilon_p(l))=\frac{1}{e^{(\epsilon_p(l)-\mu_l)/T_l}+1}\,,
\end{equation}
\begin{equation}
f^{L}(r,\epsilon_p(r))=\frac{1}{e^{(\epsilon_p(r)-\mu_r)/T_r}+1}\,,
\end{equation}
\end{subequations}
where $\mu_l=\mu+eV$, $\mu_r=\mu$ and $T_l=T+\Delta T$, $T_r=T$. [We
use shorthand notation for the distribution function of right movers
at the left lead $f^{R}(l,\epsilon_p(l))=f^{R}(x=0,\epsilon_p(x=0))$
and left movers at the right lead
$f^{L}(r,\epsilon_p(r))=f^{L}(x=L,\epsilon_p(x=L))$.] The
parametrization, Eq.~\eqref{f=fR+fL}, is especially useful since
owing to the simple algebraic relation,
\begin{equation}
v_p\frac{\partial f^{R(L)}}{\partial\epsilon}\frac{\partial
\epsilon}{\partial x}-\frac{\partial U}{\partial x}\frac{\partial
f^{R(L)}}{\partial \epsilon}\frac{\partial \epsilon}{\partial p}=0
\end{equation}
the inhomogeneity-related term drops out from the left-hand side of
the kinetic equation, except for the residual contribution
$\delta(p)\partial_x U(x)[f^{R}(x,U(x))-f^{L}(x,U(x))]$ at $p=0$.
For noninteracting electrons the mismatch between distribution
functions $f^{R/L}(x,U(x))$ of right and left movers is
exponentially small at the bottom of the band.  In addition, even
this small discontinuity is smeared by inter-electron scattering
responsible for equilibration.\cite{Tobias} It is thus safe to take
$f^R(x,\epsilon_p(x))=f^{L}(x,\epsilon_p(x))$ for $p=0$ and we get
then instead of Eq.~\eqref{KE},
\begin{equation}\label{KE-fR-fL}
\theta(p)v_p\partial_{x}f^{R}(x,\epsilon_{p}(x))
+\theta(-p)v_p\partial_{x}f^{L}(x,\epsilon_{p}(x))=\mathcal{I}\{f\}\,.
\end{equation}

As the first step of our general analysis, we demonstrate now with
the help of kinetic Eq.~\eqref{KE-fR-fL} that deviations in electric
and thermal conductances from their noninteracting values
[Eqs.~\eqref{G-0} and \eqref{K-0}] are ultimately related to the
rate of change in the number of say right-moving electrons
$\dot{N}^R$ and heat exchange rate $\dot{Q}^R$ between right movers
and left movers.

%------------------------------------------------------------------------------------------------------------
\subsection{Conservation laws}\label{Sec-Formalism-ConservaionLaws}

The rate of change in the number of right movers $\dot{N}^R$ due to
electron collisions is obtained from the collision integral
$\mathcal{I}\{f\}$ upon integration over positive momenta and wire
length,
\begin{equation}
\dot{N}^{R}=\frac{2}{h}\int^{L}_{0} dx \int^{\infty}_{0}dp\,
\mathcal{I}\{f\}\,,
\end{equation}
where the coefficient 2 stands for two spin projections. Owing to
the Boltzmann Eq.~\eqref{KE-fR-fL} $\dot{N}^R$ can be equivalently
presented in terms of the distribution function of right-moving
electrons as
\begin{equation}
\dot{N}^{R}=\frac{2}{h}\int^{L}_{0}
dx\int^{\infty}_{U(x)}d\epsilon\,\partial_x f^{R}(x,\epsilon)\,.
\end{equation}
We can integrate this expression by parts by noticing that
\begin{equation}\label{Int-by-parts}
\frac{\partial}{\partial x}\!\int^{\infty}_{U(x)}\!\!\!d\epsilon
f^R(x,\epsilon)=\int^{\infty}_{U(x)}\!\!\!d\epsilon\,
\partial_{x}f^{R}(x,\epsilon)-f^{R}(x,U)\partial_x U(x),
\end{equation}
and approximating in the following the distribution function of
right movers by unity at the bottom of the band, which is correct up
to exponentially small terms at low temperatures
$f^{R}(x,U)\approx1-\mathcal{O}[e^{-(\mu-U)/T}]$. This would give
then
\begin{equation}\label{NR-U}
\dot{N}^{R}=j^{R}(r)-j^{R}(l)+\frac{2}{h}[U(r)-U(l)]\,,
\end{equation}
where we used standard definition for the currents of right/left
movers,
\begin{equation}\label{j}
j^{R/L}(x)=\frac{2}{h}\int^{+\infty}_{-\infty}\! dp\,\,\theta(\pm
p)v_p f(x,\epsilon_p)\,.
\end{equation}
In Eq.~\eqref{NR-U} the incoming current $j^R(l)$ of right movers at
$x=0$ is known since it is controlled by the distribution of
noninteracting electrons in the left lead,
Eq.~\eqref{f-RL-boundary}. However, the outgoing $j^R(r)$ is not
known because the distribution function of right-movers varies along
the wire as a result of scattering.  It is convenient to exclude
this unknown from Eq.~\eqref{NR-U} by noticing that the total
current $j$ in the wire does not depend on position due to the
conservation of the number of electrons and can be written as
$j=j^R(r)+j^{L}(r)$.  This results in
\begin{equation}\label{NR-U-1}
\dot{N}^{R}=j-[j^{L}(r)+j^{R}(l)]+\frac{2}{h}[U(r)-U(l)]\,.
\end{equation}
The benefit of performing this step is that now both currents
$j^{R}(l)$ and $j^{L}(r)$ are controlled by the noninteracting leads
whose distribution functions are given by the boundary conditions
[Eq.~\eqref{f-RL-boundary}].  Furthermore, since $\dot{N}^{R}$ and
$j$ vanish in the absence of applied bias, we can exclude the
$U$-dependent contribution from Eq.~\eqref{NR-U-1} by subtracting
from it $j^{L}(r)+j^{R}(l)|^{\Delta
  T=0}_{V=0}=\frac{2}{h}[U(r)-U(l)]$. This leads to
\begin{equation}\label{NR-statistical}
\dot{N}^R=j-\left[j^R(l)-j^{R}(l)|^{\Delta T=0}_{V=0}\right]\,.
\end{equation}
The difference between currents of right movers at the left boundary
with and without bias in Eq.~\eqref{NR-statistical} can be found
with the help of distribution function \eqref{f-RL-boundary}.
Indeed, after a simple calculation
\begin{eqnarray*}
&&\hskip-.5cm j^{R}(l)-j^{R}(l)|^{\Delta T=0}_{V=0}
\\
&&\hskip-.5cm
=\frac{2}{h}\int^{\infty}_{0}\!dp\,v_p\left[f^{R}(l,\epsilon_{p}(l))-
f^{R}(l,\epsilon_{p}(l))|^{\Delta
T=0}_{V=0}\right]=\frac{2eV}{h},\nonumber
\end{eqnarray*}
valid up to corrections small as $e^{-\mu/T}$, we find
\begin{equation}\label{Int-Landauer}
\frac{2e^2}{h}V=I-e\dot{N}^R\,,
\end{equation}
where $I=ej$. This result can be thought of as a generalization of
Landauer formula for interacting one-dimensional systems. In the
noninteracting limit $\dot{N}^R=0$ and we recover $G=I/V=2e^2/h$
while a finite $\dot{N}^R$ would lead to a change in the
conductance. Equation \eqref{Int-Landauer} was derived earlier for
uniform (clean) wires.~\cite{Jerome-3,Tobias} We have shown here
that it remains intact even in the case of inhomogeneous wires.

We now repeat the above calculation for the energy change of
right-movers $\dot{E}^R$ induced by electron collisions. The latter
is obtained from the collision integral $\mathcal{I}\{f\}$ by
multiplying it by $\epsilon_p$ and then integrating over positive
momenta and the wire length
\begin{equation}
\dot{E}^{R}=\frac{2}{h}\int^{L}_{0}dx\int^{\infty}_{0}dp\,\epsilon_p
\mathcal{I}\{f\}\,.
\end{equation}
It can be equivalently rewritten in terms of $f^{R}(x,\epsilon)$ by
making use of the Boltzmann Eq.~\eqref{KE-fR-fL},
\begin{equation}
\dot{E}^{R}=\frac{2}{h}
\int^{L}_{0}dx\int^{\infty}_{U(x)}d\epsilon\,\epsilon\,
\partial_x f^{R}(x,\epsilon)\,.
\end{equation}
After integration by parts, similar to Eq.~\eqref{Int-by-parts}, one
finds
\begin{equation}\label{ER-U}
\dot{E}^R=j^{R}_{E}(r)-j^{R}_{E}(l)+\frac{1}{h}\left[U^2(r)-U^{2}(l)\right]\,,
\end{equation}
where we used the usual definition for energy currents of right/left
movers,
\begin{equation}\label{jE}
j^{R/L}_E(x)=\frac{2}{h}\int^{+\infty}_{-\infty} dp\,\theta(\pm
p)v_p\epsilon_p f(x,\epsilon_p)\,.
\end{equation}
Conservation of energy ensures that the energy current $j_E$ is
constant along the wire. By using it at the right end
$j_E=j^{R}_{E}(r)+j^{L}_{E}(r)$, we can exclude unknown
$j^{R}_{E}(r)$ from Eq.~\eqref{ER-U} in analogy with
Eq.~\eqref{NR-U-1}. In addition, since $\dot{E}^R=0$ and $j_E=0$
without the bias, we subtract $j^{L}_{E}(r)+j^{R}_{E}(l)|^{\Delta
T=0}_{V=0}=\frac{1}{h}[U^2(r)-U^2(l)]$ to exclude the $U$-dependent
contribution. This procedure gives for the rate of energy change
\begin{equation}\label{ER-statistical}
\dot{E}^R=j_{E}-\left[j^R_{E}(l)-j^{R}_{E}(l)|^{\Delta
T=0}_{V=0}\right]\,.
\end{equation}
The energy current of right movers at the left end of the wire is
controlled by the noninteracting lead with known distribution
function \eqref{f-RL-boundary}. To linear order in $V$ and $\Delta
T$ a simple calculation gives us
\begin{eqnarray}
&&j^{R}_E(l)-j^{R}_E(l)|^{\Delta T=0}_{V=0}\nonumber
\\
&&=\frac{2}{h}\int^{\infty}_{0}dp\,v_p\epsilon_p\left[f^{R}(l,\epsilon_{p}(l))-
f^{R}(l,\epsilon_{p}(l))|^{\Delta T=0}_{V=0}\right]\nonumber
\\
&&=\frac{2eV\mu}{h}+\frac{2\pi^2T\Delta T}{3h}\,.
\end{eqnarray} It is
convenient to combine electric and energy currents into the heat
current
\begin{equation}
j_Q=j_E-\mu j\,,
\end{equation}
for which we find from Eqs.~\eqref{Int-Landauer} and
\eqref{ER-statistical},
\begin{equation}\label{Heat-balance}
\frac{2\pi^2 }{3h}T\Delta T=j_Q-\dot{Q}^R\,,
\end{equation}
where $\dot{Q}^R=\dot{E}^R-\mu\dot{N}^R$ is the heat transferred
into the right-moving subsystem by electron collisions. As expected
when $\dot{Q}^R=0$ we recover from Eq.~\eqref{Heat-balance} the
thermal conductance of noninteracting electrons Eq.~\eqref{K-0},
while a nonvanishing $\dot{Q}^R$ results in an interaction-induced
change in $K$. Equation \eqref{Heat-balance} was reported earlier
for uniform wires~\cite{Jerome-3,Tobias} and as it is shown here
remains valid even in the inhomogeneous case.

So far our basic Eqs.~\eqref{Int-Landauer} and \eqref{Heat-balance}
contain four unknown entries: two response currents $I$ and $j_Q$
due to applied bias $V$ and temperature difference $\Delta T$, and
two rates $\dot{N}^R$ and $\dot{Q}^R$ nonvanishing due to
interactions. In the next sections we demonstrate that all four
quantities can be expressed in terms of only two parameters $u(x)$
and $\Delta\mu(x)=\mu^R(x)-\mu^L(x)$ which enter the distribution
function \eqref{f-part-eq} of partially equilibrated electrons. This
would give us the closed set of equations that relate
$\{I,j_Q\}\rightleftarrows\{V,\Delta T\}$ and thus determine the
transport coefficients of interest.

%---------------------------------------------------------------------------------------------------------
\subsection{Currents $I$ and $j_Q$ in the partially equilibrated wires}\label{Sec-Formalism-Ij}

Electric and heat currents can be easily found knowing the
distribution function $f(x,p)$ of electrons in the wire. As we
discussed in Sec.~\ref{Sec-UnifromWire} for the partially
equilibrated wire this distribution is given by
Eq.~\eqref{f-part-eq}. It is worth emphasizing that for the
inhomogeneous case all four parameters entering
Eq.~\eqref{f-part-eq}: velocity $u$, chemical potential of right-
and left-moving electrons $\mu^{R(L)}$, and effective temperature
$\bar{T}$, are, in principle, coordinate dependent. Furthermore, the
distribution, Eq.~\eqref{f-part-eq}, does not apply to particles
near the bottom of the band, for $|p|\lesssim\sqrt{mT}$, as
explained later in the text (see also corresponding discussions in
Ref.~\onlinecite{Tobias}). This, however, does not cause any extra
difficulties since transport quantities of interest are determined
by the behavior of the distribution function near the Fermi level.

By using the definition of the current, Eq.~\eqref{j}, and
distribution function \eqref{f-part-eq} we obtain the electric
current in the partially equilibrated wire
\begin{equation}\label{I}
I=\frac{2e}{h}\Delta\mu(x)+en(x)u(x)\,.
\end{equation}
Notice here that although contributions to the current due to the
electron drift, $en(x)u(x)$, and partial equilibration between right
and left movers, $2e\Delta\mu(x)/h$, are individually coordinate
dependent, their sum must be constant along the wire. This is a
consequence of the particle number conservation.

Since the heat current $j_Q$ also does not depend on position, it
can be calculated at any point in the wire. In the regions not too
close to the leads the distribution function is expected to have the
partially equilibrated form \eqref{f-part-eq}. Then using
expressions \eqref{j} and $\eqref{jE}$ for $j$ and $j_E$ we obtain
after Sommerfeld expansion of the integrand to leading order in
$T/\mu\ll1$,
\begin{equation}\label{jQ}
j_Q=\frac{\pi^2}{6}\frac{T^2}{\mu(x)}n(x)u(x)\,,
\end{equation}
where we introduced $\mu(x)=\mu-U(x)$. Since $j_Q$ is already
proportional to small $u\propto V$, we were able to replace
$\bar{T}$ with $T$ within the linear-response regime.  Notice also
that the particular combination $n(x)u(x)/\mu(x)$ defining heat
current $j_Q$ must be coordinate independent.

To make further progress we should elaborate on the expressions for
the rates $\dot{N}^R$ and $\dot{Q}^R$, whose explicit forms depend
on details of the equilibration mechanism. This can be done
following the idea suggested in Ref.~\onlinecite{Tobias} and we show
below how $\dot{N}^R$ and $\dot{Q}^R$ can be expressed through
$\Delta\mu(x)$ and $u(x)$.

\begin{figure}
  \includegraphics[width=8cm]{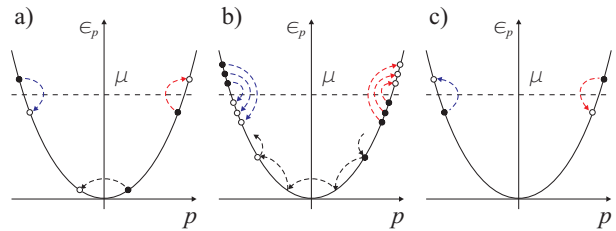}
  \caption{[Color online] (a) Dominant three-particle collision which
  changes the number of right-moving electrons. (b) Equilibration mechanism, multi-step
  backscattering of right mover into the left mover. At low
  temperatures each step $\delta p$ in momentum space is of the order of $\sim
  T/v_F$.
  (c) Energy-conserving two-particle scattering process that violates conservation of
  momentum. This process is possible due to the presence of inhomogeneities.}\label{Fig-3pc}
\end{figure}

%---------------------------------------------------------------------------------------------------------
\subsection{Microscopic expressions for $\dot{N}^R$ and $\dot{Q}^R$}\label{Sec-Formalism-NRQR}

First let us identify the leading backscattering mechanism that
contributes to $\dot{N}^R$. The most favorable collisions should
involve a maximal number of electronic states close to the Fermi
points. However, due to the conservation of total energy and
momentum, collisions that change the number of right and left movers
cannot occur near the Fermi level only, and have to involve states
deep in the electron band. As was pointed out in
Ref.~\onlinecite{Lunde} the scattering process most important in
altering the current thus typically scatters two electrons close to
the Fermi points and one electron at the bottom of the band, as
schematically depicted in Fig.~\ref{Fig-3pc}a. It is convenient to
think of this collision as a process in which a deep hole,
corresponding to the outgoing electron state, is backscattered by
electron excitations close to the Fermi level. These excitations are
typically associated with a momentum change $\delta p\sim T/v_F\ll
p_F$ due to Fermi blocking. Let us furthermore characterize this
process by introducing three-particle scattering rate
$1/\tau_{eee}$, which can be approximated by a constant because the
initial and final states both lie at the bottom of the band. Since
the sign of $\delta p$ varies in a random fashion from one collision
to another the hole performs a Brownian motion in momentum space.
The corresponding diffusion coefficient $B$ can be readily
estimated. The typical momentum change of a hole over time $t$
behaves as $(\Delta p)^2\sim Bt$. As we assumed the hole changes its
momentum by $\pm T/v_F$ once during the time $\tau_{eee}$, so we
conclude that $(\Delta p)^2\sim(T/v_F)^2t/\tau_{eee}$ for
$t\gg\tau_{eee}$ and thus estimate
\begin{equation}
B\sim \frac{T^2}{v^2_{F}\tau_{eee}}\,.
\end{equation}

The change $\Delta \dot{N}^R$ in the number of right-moving
electrons over the time $t\sim(\Delta p)^2/B$ for the segment of the
wire $\Delta x$ is given by the rate $t^{-1}$ times the number of
deep holes susceptible to be backscattered. The latter can be
estimated from the probability to find a left- or right-moving hole
$e^{-\mu^{L(R)}/T}$ and the number of states $\Delta p\Delta x/h$
available within the typical momentum range $\Delta p\sim\sqrt{mT}$
of the backscattering processes shown in Fig.~\ref{Fig-3pc}b. Taking
into account that the scattering of left- and right-moving holes
both contribute to $\dot{N}^R$, but with opposite signs, one finally
estimates
\begin{eqnarray}
\Delta \dot{N}^R\sim\frac{1}{t}\left(\frac{\Delta p\Delta
x}{h}\right)\left(e^{-\mu^R/T}-e^{-\mu^L/T}\right)\nonumber\\
\approx-\frac{\Delta\mu\Delta x B}{h\sqrt{mT^3}}e^{-\mu/T}\,,
\end{eqnarray}
with $\Delta\mu=\mu^R-\mu^L$.  A careful calculation based on the
kinetic equation gives~\cite{Tobias}
\begin{equation}\label{NR-kinetic}
\frac{d\dot{N}^R}{dx}=-\frac{2\Delta\mu(x)}{h}\frac{e^{-\mu(x)/T}}{\ell_1(x)}\,,\quad
\ell_{1}(x)=\frac{\sqrt{8\pi mT^3}}{B(x)}\,.
\end{equation}
The expression for the diffusion coefficient $B$ is model specific.
Our preliminary estimate gives $B\propto T^3$ in the case of Coulomb
interaction.

We continue now with the calculation of the rate $\dot{Q}^R$, which
consists of two contributions
\begin{equation}\label{QRb+QRp}
\dot{Q}^{R}=\dot{Q}^{R}_{b}+\dot{Q}^{R}_p\,.
\end{equation}
The first one $\dot{Q}^{R}_{b}$ is related to the same
backscattering events that change the number of right-movers
$\dot{N}^{R}$. Exploring the fact that both rates are caused by the
same physical processes it was shown in Refs.~\onlinecite{Jerome-3}
and \onlinecite{Tobias} that there is a relation between $\dot{N}^R$
and $\dot{Q}^R_b$, which we generalize here for the inhomogeneous
case,
\begin{equation}\label{QRb}
\frac{d\dot{Q}^{R}_{b}}{dx}=-2\mu(x)\frac{d\dot{N}^R}{dx}\,.
\end{equation}
The logic behind this equation is as follows. The backscattering
processes transform the unperturbed distribution of electrons into
the partially equilibrated form \eqref{f-part-eq} with nonvanishing
$u(x)$. The two distributions differ most prominently at energies
within $\sim T$ of the Fermi level. One can thus assume that all the
right-moving electrons contributing to $\dot{N}^R$ are removed from
the vicinity of the right Fermi point and placed to the vicinity of
the left one. Each such transfer reduces the momentum of the system
by $2p_F$. The other electrons have to be scattered in the
vicinities of the two Fermi points to accommodate this momentum
change. In the special case of three-particle collisions, the
transfer of electron from the right Fermi point to the left one is
accomplished in a number of small steps with momentum change $\delta
p\sim T/v_F$, and at each step one additional electron is scattered
near each of the two Fermi points, see Fig.~\ref{Fig-3pc}b. As a
result of the rearrangement of electrons near the two Fermi points,
the local momentum change $2p_F$ of the backscattered electrons is
distributed between the remaining right- and left-moving electrons,
i.e., $\delta p^R+\delta p^L=2p_F$. Thus the energy of the remaining
right movers increases by $\delta E^R =v_F\delta p^R$ whereas that
of the left movers decreases, $\delta E^L=-v_F\delta p^L$. Then, the
conservation of energy requires $\delta p^R=\delta p^L=p_F$. In the
end, the energy balance for the right-moving electrons consists of a
loss of $\mu$ due to the removal of one particle from the Fermi
level and a gain of $\delta E^R=v_Fp_F=2\mu$ due to the
redistribution of momentum. As a result, for every right-moving
electron that changes direction, $\Delta N^R=-1$, the right-movers
energy increases by an amount $\Delta E^R=\mu$, so one concludes
that $\dot{E}^R=-\mu\dot{N}^R$ or equivalently
$\dot{Q}^R=-2\mu\dot{N}^R$. Equation \eqref{QRb} follows from here
naturally if one applies the same argument but for the segment of
wire $\Delta x$ such that rates $\dot{N}^R$ and $\dot{Q}^R$ are
accounted per unit of length in the inhomogeneous wire.

The other contribution $\dot{Q}^{R}_{p}$ to the heat transferred by
right movers in Eq.~\eqref{QRb+QRp} is due to scattering processes
that do not conserve momentum, see Fig.~\ref{Fig-3pc}c for
illustration. These two-body collisions are possible only in the
inhomogeneous case. They do not change the number of right-moving
electrons, but do change their energy. It is expected that this rate
is proportional to the velocity $u$ of the electron liquid,
$\dot{Q}^R_p\propto u$. Indeed, two-particle collisions of
Fig.~\ref{Fig-3pc}c involve a right mover with momentum $p\approx
p_F$ and a left mover with momentum $p\approx-p_F$. For these
electrons the drift term $pu\approx\pm p_Fu$ in the distribution
function \eqref{f-part-eq} can be absorbed into the temperature
$\bar{T}$, such that right movers can be considered as being at an
effective temperature $T^R\approx\bar{T}(1+u/v_F)$ while left movers
at temperature $T^L\approx\bar{T}(1-u/v_F)$, to linear order in
$u$.~\cite{Jerome-2} According to the general principle of
statistical mechanics thermalization between these subsystems
involves the energy flow from ``warmer'' right movers to ``colder''
left movers that is proportional to the difference in temperatures
between the two, $\dot{Q}^R_p\propto T^R-T^L\propto u$. An explicit
microscopic calculation of the rate $\dot{Q}^R_p$ done in
Appendix~(\ref{Sec-Appendix-QR}) gives
\begin{equation}\label{QRp}
\frac{d\dot{Q}^R_p}{dx}=-2\mu(x)\frac{n(x)u(x)}{\ell_{\mathrm{in}}(x)}\,.
\end{equation}
Here $\ell_{\mathrm{in}}$ is a scattering length scale associated
with these momentum-nonconserving collisions,  Fig.~\ref{Fig-3pc}c,
\begin{equation}\label{l-in}
\ell^{-1}_{\mathrm{in}}=\frac{\Upsilon(x)}{16n(x)}\frac{T}{\mu(x)}\,,
\end{equation}
where the parameter
\begin{eqnarray}\label{l-gamma}
&&\hskip-.75cm\Upsilon(x)=\left\{\left[\partial_{x}\!
\left(\frac{\mathcal{V}_0-\mathcal{V}_{2k_F(x)}}{\pi\hbar
v_F(x)}\right)\right]^2\right.\nonumber\\
&&\hskip-.75cm\left.+\left[\partial_{x}\!
\left(\frac{\mathcal{V}_0}{\pi\hbar v_F(x)}\right)\right]^2\!\!\!
+\left[\partial_{x}\! \left(\frac{\mathcal{V}_{2k_F(x)}}{\pi\hbar
v_F(x)}\right)\right]^2 \right\}
\end{eqnarray}
is expressed through the zero momentum and $2k_F$ Fourier components
of the electronic interaction potential $\mathcal{V}$. The complete
rate $\dot{Q}^R$ is thus given by the sum of Eqs.~\eqref{QRb} and
\eqref{QRp} and we find
\begin{equation}\label{QR-kinetic}
\frac{d\dot{Q}^R}{dx}=-2\mu(x)\frac{d\dot{N}^R}{dx}-
2\mu(x)\frac{n(x)u(x)}{\ell_{\mathrm{in}}(x)}\,.
\end{equation}

One should make two important comments regarding this result. First,
the effect of inhomogeneity on resistivity of a quantum wire was
recently addressed in Refs.~\onlinecite{Jerome-1,Jerome-2} assuming
that electrons are fully equilibrated and thus described by the
distribution function \eqref{f-eq}. This assumption requires that
three-body interaction processes which change the number of
right-moving electrons dominate over the momentum-nonconserving
scattering. However, in a situation where both interaction processes
happen on a comparable time scale the system is frustrated with a
finite value of $\Delta\mu\neq0$ and thus $u\neq v_d=I/en$, so that
electrons are described by the distribution, Eq.~\eqref{f-part-eq},
which we used in our calculations.

To elucidate further the origin of the frustration it is important
to emphasize that momentum-nonconserving scattering resists full
equilibration of electrons into a single distribution,
Eq.~\eqref{f-eq}. These scattering processes reduce velocity $u$ and
thus prevent complete relaxation of the difference in chemical
potentials $\Delta\mu$. Indeed, since the current $I$ is fixed by
the external circuit then according to Eq.~\eqref{I} decrease in $u$
implies increase in $\Delta\mu$. This effect is opposite to that of
the equilibration processes due to three-particle collisions which
tend to relax $\Delta\mu$.

Second, the correction to resistance of a nonuniform wire was
obtained in studies~\cite{Jerome-1,Jerome-2} by calculating the rate
of momentum change $\dot{P}^R$ for right-moving electrons. Note,
however, that in an inhomogeneous system without translational
invariance momentum is not a good quantum number. Indeed, the
momentum change of right movers due to collisions depends on
position and thus, is not the same for electrons inside the wire,
where it was calculated,~\cite{Jerome-1,Jerome-2} than that (actual
change) in the leads. Our present scheme is free from this
difficulty. As we show in the next section it is really the rate
$\dot{Q}^R$, computed from the well-defined energy exchange, that is
needed to determine the wire resistance and other transport
coefficients.

%------------------------------------------------------------------------------------------------------------
\section{Transport coefficients}\label{Sec-Transport}

We are set now for the calculation of transport coefficients in an
inhomogeneous wire. Indeed, our basic Eqs.~\eqref{Int-Landauer} and
\eqref{Heat-balance} provide electric $I$ and heat $j_Q$ currents as
the response to the applied bias $V$ and temperature difference
$\Delta T$. Interaction effects are captured by the rates
$\dot{N}^R$ and $\dot{Q}^R$ induced by particle collisions that
encode change in the number of right-moving electrons and energy
exchange between right- and left-movers, respectively. These rates
are defined by Eqs.~\eqref{NR-kinetic} and $\eqref{QR-kinetic}$
which still contain the unknown difference between the chemical
potentials of partially equilibrated right and left movers
$\Delta\mu(x)$ and flow velocity $u(x)$. Equations~\eqref{I} and
\eqref{jQ} are the final ingredients that allow to establish the
correspondence $\{V,\Delta T\}\rightleftarrows\{I,j_Q\}$ and thus
find transport coefficients of interest.

Technically one proceeds as follows. First, employing conservation
of currents (recall that $I$ and $j_Q$ are constant along the wire)
one can express $\Delta\mu(x)$ and $u(x)$ in terms of $I$ and $j_Q$
from Eqs.~\eqref{I} and \eqref{jQ}. Second, one brings these
relations into Eqs.~\eqref{NR-kinetic} and $\eqref{QR-kinetic}$ to
find the rates $\dot{N}^R$ and $\dot{Q}^R$ in terms of $I$ and
$j_Q$, which is possible to do in quadratures. Finally,
Eqs.~\eqref{Int-Landauer} and \eqref{Heat-balance} define the
desired correspondence $\{V,\Delta T\}\rightleftarrows\{I,j_Q\}$.
Employing this procedure we find two independent linear equations,
\begin{eqnarray}
&&\hskip-.75cm
\frac{2e^2}{h}V=I(1+r_1)-\frac{6}{\pi^2}ej_Qr_1\frac{\overline{\mu}}{T^2}\,,\label{Eq1}
\\
&&\hskip-.75cm \frac{2\pi^2e}{3h}T\Delta
T=ej_Q\left[1+\frac{12}{\pi^2}
\frac{r_1\overline{\mu^2}+\gamma^2}{T^2}\right]-
2Ir_1\overline{\mu}\,,\label{Eq2}
\end{eqnarray}
where we introduced the dimensionless parameter,
\begin{equation}
r_1=\int^{L}_{0}\frac{dx}{\ell_{1}(x)}e^{-\mu(x)/T}\,,\label{r1-average}
\end{equation}
which quantifies the rate of three-particle processes that change
number of right-moving electrons [see Eq.~\eqref{NR-kinetic}], as
well as the weighted chemical potentials along the wire,
\begin{eqnarray}
&&\overline{\mu}=\frac{1}{r_1}
\int^{L}_{0}\frac{dx}{\ell_{1}(x)}\mu(x)e^{-\mu(x)/T}\,,\label{mu-average}\quad\\
&&\overline{\mu^2}=\frac{1}{r_1}
\int^{L}_{0}\frac{dx}{\ell_{1}(x)}\mu^2(x)e^{-\mu(x)/T}\,,\label{mu2-average}\\
&&\gamma^2=
\int^{L}_{0}\frac{dx}{\ell_{\mathrm{in}}(x)}\mu^{2}(x)\label{mu-gamma-average}\,.
\end{eqnarray}
From Eqs.~\eqref{Eq1} and \eqref{Eq2} we find the resistance
\begin{equation}\label{R}
R=\left.\frac{V}{I}\right|_{\Delta T=0}=\frac{h}{2e^2}[1+r]\,,
\end{equation}
where
\begin{equation}\label{R-correction}
r=r_1-\frac{r^2_1\overline{\mu}^2}{\frac{\pi^2}{12}T^2+r_{1}\overline{\mu^2}
+\gamma^2}\,,
\end{equation}
and Peltier coefficient
\begin{equation}\label{Pi}
\Pi=\left.\frac{j_Q}{I}\right|_{\Delta
T=0}=\frac{\pi^2T^2}{6e}\frac{r_1\overline{\mu}}{\frac{\pi^2}{12}T^2+r_{1}\overline{\mu^2}
+\gamma^2}\,.
\end{equation}
In addition we find the thermal conductance
\begin{equation}\label{K}
K=\left.\frac{j_Q}{\Delta
T}\right|_{I=0}=\frac{\pi^4T^3}{18h}\frac{1}{\frac{\pi^2}{12}T^2+r_{1}\overline{\mu^2}
+\gamma^2}\,,
\end{equation}
and thermopower
\begin{equation}\label{S}
S=-\left.\frac{V}{\Delta
T}\right|_{I=0}=\frac{\pi^2T}{6e}\frac{r_1\overline{\mu}}
{\frac{\pi^2}{12}T^2+r_{1}\overline{\mu^2}+\gamma^2}\,.
\end{equation}
Predictably, the Peltier coefficient, Eq.~\eqref{Pi}, and
thermopower, Eq.~\eqref{S}, satisfy the Onsager relation $\Pi=ST$.
Equations \eqref{R}--\eqref{S} are the main results of this paper.
In the following we analyze these general expressions for a few
modeling examples of inhomogeneities.

%------------------------------------------------------------------------------------------------------------
\subsection{Uniform wire}

First of all, we perform a consistency check for the case of a
uniform wire, recently studied in Ref.~\onlinecite{Tobias}. In the
homogeneous case $U(x)\to0$ all quantities defining $R$, $\Pi$, $K$
and $S$ become coordinate independent: $\ell_{1}(x)\to\ell_{1}$ so
that $\overline{\mu^2}\equiv\overline{\mu}^2=\mu^2$ and
$r_1\to(L/\ell_1)e^{-\mu/T}$. At the same time $\gamma\to0$ which is
a consequence of momentum conservation: in a uniform wire
two-electron scattering processes, shown in Fig.~\ref{Fig-3pc}c, are
not allowed. As a result, interaction-induced correction to the wire
resistance, Eq.~\eqref{R-correction}, reduces to
\begin{equation}\label{R-unifrom}
r=\frac{r_0r_1}{r_0+r_1}\,,
\end{equation}
where $r_{0}=\pi^2T^2/12\mu^2$. In order to establish a connection
with the notations of Ref.~\onlinecite{Tobias} we invert resistance,
Eq.~\eqref{R}, with $r$ taken from Eq.~\eqref{R-correction}, to get
the conductance $G=R^{-1}$ to leading order in $T/\mu\ll1$ and find
\begin{equation}\label{G-uniform}
G=\frac{2e^2}{h}\left[1-\frac{\pi^2}{12}\frac{T^2}{\mu^2}\frac{L}{L+\ell_{\mathrm{eq}}}\right]\,,
\end{equation}
where following Ref.~\onlinecite{Tobias} we have introduced the
equilibration length
\begin{equation}\label{l-eq}
\ell_{\mathrm{eq}}=\frac{\pi^2}{12}\frac{T^2}{\mu^2}\ell_1e^{\mu/T}\,.
\end{equation}
This result shows that for a long wire $L\gg\ell_{\mathrm{eq}}$ the
conductance saturates to its length independent value, which still
exhibits noticeable power-law correction in temperature, $\delta
G=-(2e^2/h)(\pi^2T^2/12\mu^2)$, already mentioned in
Sec.~\ref{Sec-UnifromWire} [Eq.~\eqref{G-eq}]. This saturation of
conductance is expected, since the electronic system reaches full
equilibrium.  For short wires, $\ell_1\ll L\ll \ell_{\mathrm{eq}}$,
the interaction-induced correction to conductance is exponentially
small $\delta G=-(2e^2/h)(L/\ell_1)e^{-\mu/T}$ and scales linearly
with the length of the wire.  It is worth noting that
Eq.~\eqref{G-uniform} is only applicable to wires longer than
$\ell_1$. This constraint comes from the approximations made when
deriving the rate of change for the right-moving electrons
$\dot{N}^R$ in Eq.~\eqref{NR-kinetic} [see Ref.~\onlinecite{Tobias}
for details].

The Peltier coefficient and thermopower of a uniform wire follow
from Eqs.~\eqref{Pi} and \eqref{S}, and read
\begin{equation}\label{Pi-uniform}
\Pi=ST=\frac{\pi^2}{6e}\frac{T^2}{\mu}\frac{L}{L+\ell_{\mathrm{eq}}}\,.
\end{equation}
It shows that $\Pi$ grows from exponentially small values at
$L\ll\ell_{\mathrm{eq}}$ to $\Pi_{\mathrm{eq}}$ quoted in
Eq.~\eqref{Pi-S-eq} at $L\gg\ell_{\mathrm{eq}}$. The thermal
conductance behaves very differently though. One finds from
Eq.~\eqref{K} in the uniform limit
\begin{equation}\label{K-uniform}
K=\frac{2\pi^2T}{3h}\frac{\ell_{\mathrm{eq}}}{L+\ell_{\mathrm{eq}}}\,.
\end{equation}
At $L\ll\ell_{\mathrm{eq}}$ one recovers the result,
Eq.~\eqref{K-0}, for noninteracting wires, but as the length of the
wire grows, $K$ is suppressed as $1/L$ and vanishes for fully
equilibrated wires, see our earlier discussion presented below
Eq.~\eqref{K-eq}. Equations \eqref{Pi-uniform} and \eqref{K-uniform}
recover the corresponding results of Ref.~\onlinecite{Tobias}.

%------------------------------------------------------------------------------------------------------------
\subsection{Two wires in series}\label{Sec-Transport-TwoWires}

As the simplest prototype of nonuniform system we study two uniform
wires of lengths $L_1$ and $L_2$, with different densities,
connected in series with each other. We ignore here small
contribution to transport coefficients coming from $\gamma$, which
is nonzero only in the near vicinity of the junction between the
wires and vanishes everywhere else. Applying then
Eqs.~\eqref{mu-average}--\eqref{mu-gamma-average} to this setup we
find for the interaction-induced resistance,
Eq.~\eqref{R-correction},
\begin{equation}\label{R-two-wires}
r\!=\!\frac{\pi^2T^2}{12}
\frac{\frac{L_{1}}{\mu^2_1\ell^{(1)}_{\mathrm{eq}}}+\frac{L_{2}}{\mu^2_2\ell^{(2)}_{\mathrm{eq}}}+
\left[\frac{1}{\mu_1}-\frac{1}{\mu_2}\right]^2
\frac{L_{1}L_{2}}{\ell^{(1)}_{\mathrm{eq}}\ell^{(2)}_{\mathrm{eq}}}}
{1+\frac{L_1}{\ell^{(1)}_{\mathrm{eq}}}+\frac{L_2}{\ell^{(2)}_{\mathrm{eq}}}}\,,
\end{equation}
with chemical potentials $\mu_{i}=\mu-U_{i}$ and equilibration
lengths
$\ell^{(i)}_{\mathrm{eq}}=(\pi^2T^2/12\mu^2_{i})\ell^{(i)}_{1}e^{\mu_{i}/T}$
within each wire $i=1,2$. It is of special interest to consider the
limit when one wire is infinitely long. Upon taking the limit
$L_1\to\infty$ the expression for $r$ reduces to
\begin{equation}\label{R-two-wires-lim}
r\to\frac{\pi^2T^2}{12\mu^{2}_{1}}+\frac{\pi^2}{12}
\left[\frac{T}{\mu_1}-\frac{T}{\mu_2}\right]^2\frac{L_2}{\ell^{(2)}_{\mathrm{eq}}}\,.
\end{equation}
The first term of this formula corresponds to the residual
resistance of the first wire. It is independent of its length $L_1$,
which is natural since in the limit $L_1\to\infty$ electrons reach
full equilibration and the interaction-induced resistance should
saturate.~\cite{Jerome-3} The other contribution to the resistance
in Eq.~\eqref{R-two-wires-lim} is due to the second wire, which,
however, ceases to saturate even when
$L_2\gg\ell^{(2)}_{\mathrm{eq}}$. Thus it remains proportional to
the wire length $L_2$ and could be much larger than the first term.
The absence of equilibration in the second wire is rather
counterintuitive. This result forced us to reexamine more carefully
continuity equations for electric and heat currents. We have found
that it is not possible to match both $I$ and $j_Q$ in the wires
while simultaneously imposing vanishing chemical potential
difference $\Delta\mu$ between right- and left-moving electrons.
Indeed, let us suppose that in the limit
$L_1\gg\ell^{(1)}_{\mathrm{eq}}$ and
$L_2\gg\ell^{(2)}_{\mathrm{eq}}$ both wires are fully equilibrated,
so that $\Delta\mu=0$ in each wire. The current conservation then
becomes $I=en_1v_{d1}=en_2v_{d2}$, where $v_{d1,2}$ is the drift
velocity within each wire. According to Eq.~\eqref{jQ} the heat
current in this case can be written as
$j_{Q_{1,2}}=I(\pi^2/6e)(T^2/\mu_{1,2})$ and clearly $j_{Q1}\neq
j_{Q2}$ since $\mu_1\neq\mu_2$. The resolution of this controversy
is possible only if $\Delta\mu\neq0$ at least within one wire even
though its length exceeds the corresponding equilibration length. In
Appendix~\ref{Sec-Appendix-R} we rederived Eq.~\eqref{R-two-wires}
relying on conservation laws only.

Other transport coefficients do not show dramatic changes compared
to a single uniform wire and their behavior follows expectedly as a
natural generalization of Eqs.~\eqref{Pi-uniform} and
\eqref{K-uniform}.  We find for the Peltier coefficient of two
connected uniform wires
\begin{equation}
\Pi=\frac{\pi^2T^2}{6e}\frac{\frac{L_1}{\mu_1\ell^{(1)}_{\mathrm{eq}}}+
\frac{L_2}{\mu_2\ell^{(2)}_{\mathrm{eq}}}}
{1+\frac{L_1}{\ell^{(1)}_{\mathrm{eq}}}+\frac{L_2}{\ell^{(2)}_{\mathrm{eq}}}}\,.
\end{equation}
$\Pi$ saturates to $\pi^2T^2/6e\mu_{1(2)}$ depending on which wire
is fully equilibrated. The thermal conductance is found to be
\begin{equation}
  K=\frac{2\pi^2T}{3h}\frac{1}
  {1+\frac{L_1}{\ell^{(1)}_{\mathrm{eq}}}+\frac{L_2}{\ell^{(2)}_{\mathrm{eq}}}}\,,
\end{equation}
which is natural generalization of Eq.~\eqref{K-uniform}.

%------------------------------------------------------------------------------------------------------------
\subsection{Wire with long-range disorder}

We now study more generic models of a nonuniform wire. We assume only that
disorder variations happen on the large spatial scale, $k^{-1}_{F}\ll b\ll
L$, and concentrate on the case $\gamma^2\gg T^2$. In this case the
$\pi^2T^2/12$ term in Eq.~\eqref{R-correction} can be ignored, and the
interaction-induced resistance of the wire $r$ can be written as
\begin{equation}\label{R-crossover}
r=r_1\frac{r_1\overline{\delta\mu^2}+\gamma^2}{r_1\overline{\mu}^2+\gamma^2}\,,
\end{equation}
where we introduced
$\overline{\delta\mu^2}=\overline{\mu^2}-\overline{\mu}^2$.  Expression
\eqref{R-crossover} is applicable to any realization of long-range
disorder potential.  We now apply it to two special cases which allow
simple analytical solution.

\subsubsection{Weak disorder}

First, let us assume that amplitude $U_0$ of variations in the
inhomogeneity potential along the wire is small, $U_0\ll T$. It
turns out that Eq.~\eqref{R-crossover} covers three distinct regimes
depending on the temperature. At lowest temperatures $T\ll T_1$,
where
\begin{equation}\label{T-1}
T_1\approx\frac{\mu}{2\ln\big[\frac{\mathcal{V}_0}{\hbar v_F}k_F
b\frac{\mu}{U_0}\big]}\,,
\end{equation}
three-particle equilibration processes are weak due to exponential
suppression $e^{-\mu/T}$ of the scattering near bottom of the
band.~\cite{Polyakov} In this regime $r_1\overline{\delta\mu^2}\ll
r_1\overline{\mu}^2\ll\gamma^2$ and the resistance of the wire,
Eq.~\eqref{R-crossover}, is given by $r_1$.  It then follows from
Eq.~\eqref{r1-average} that to leading order in $U_0\ll T$
\begin{equation}\label{R-1}
r=\rho_1 L,\quad \rho_1=\frac{1}{\ell_1} e^{-\mu/T}\,,\quad T\ll
T_1\,,
\end{equation}
where $\rho_1$ has the meaning of dimensionless resistivity of the
wire. It is interesting to compare this result with
Eq.~\eqref{G-uniform} obtained for a uniform wire. In the limit
$L\ll\ell_{\mathrm{eq}}$ using Eqs.~\eqref{R} and \eqref{l-eq} we
extract from Eq.~\eqref{G-uniform} the correction to resistance
$r=(L/\ell_1)e^{-\mu/T}$ which coincides with Eq.~\eqref{R-1}.
However, there is an important difference in the applicability of
this result to uniform and disordered wires. In the case of uniform
wires $r=(L/\ell_1)e^{-\mu/T}$ applies only in the short wire limit
$L\ll\ell_{\mathrm{eq}}$, or equivalently at temperatures $T\ll
T^*_1=\mu/\ln(L/\ell_1)$. For longer wires $r$ saturates to the
length-independent value $r=\pi^2T^2/12\mu^2$, and thus always
remains smaller than contact resistance, $r\ll1$. In contrast, in
the case of disordered wires the result, Eq.~\eqref{R-1}, does not
rely on the assumption that $L\ll\ell_{\mathrm{eq}}$. The crossover
temperature, Eq.~\eqref{T-1}, is controlled by disorder and does not
depend on $L$.  Thus, although the resistivity $\rho_1$ is small,
for a sufficiently long wire the total resistance $r=\rho_1L$ can be
large, $r\gg1$.

At higher temperatures $T\gg T_1$ (strong equilibration) the resistance is
given by the sum of two terms,
\begin{equation}\label{R-crossover-2}
r=r_1\frac{\overline{\delta\mu^2}}{\overline{\mu}^2}+\frac{\gamma^2}{\overline{\mu}^2}\,.
\end{equation}
The contribution of the momentum-nonconserving two-body collisions
induced by disorder, the second term in Eq.~\eqref{R-crossover-2},
dominates in the temperature regime $T_1\ll T\ll T_2$, where
\begin{equation}
T_2\approx\frac{\mu}{2\ln\big[\frac{\mathcal{V}_0}{\hbar v_F}k_F
b\big]}\,.
\end{equation}
In this case the resistance of the wire is given
by~\cite{Jerome-1,Jerome-2}
\begin{equation}\label{R-2}
r=\rho_2L,\quad
\rho_2=\frac{\langle\Upsilon\rangle}{16n}\frac{T}{\mu}\,, \quad
T_1\ll T\ll T_2.
\end{equation}
Here we used Eqs.~\eqref{l-in} and \eqref{mu-gamma-average}, set
$\overline{\mu}^2=\overline{\mu}^2=\mu^2$ to the leading order in
$U_0\ll T$, and
$\langle\ldots\rangle=\int^{L}_{0}\frac{dx}{L}(\ldots)$ implies
averaging along the wire. In Refs.~\onlinecite{Jerome-1} and
\onlinecite{Jerome-2} the same result for resistivity was derived
assuming that electrons are fully equilibrated. Here we find that in
fact applicability conditions for $\rho_2$ are more strict and
Eq.~\eqref{R-2} dominates only in the temperature range $T_1\ll T\ll
T_2$. At higher temperatures the resistance is governed by the first
term in Eq.~\eqref{R-crossover-2},
\begin{equation}\label{R-3}
r=\rho_3L,\quad \rho_3=\frac{\pi^4T^3}{18h\mu^4}\frac{\langle
\delta\mu^2\rangle}{\varkappa}\,,\quad T_2\ll T\ll \mu\,,
\end{equation}
where $\langle\delta\mu^2\rangle=\langle U^2\rangle$ since
$\mu(x)=\mu-U(x)$. In Eq.~\eqref{R-3} we introduced thermal
conductivity of a wire $\varkappa=(2\pi^2 T/3h)\ell_{\mathrm{eq}}$.
Indeed, according to Eq.~\eqref{K-uniform} for long wires,
$L\gg\ell_{\mathrm{eq}}$, we have $K=\varkappa/L$.

Equation \eqref{R-3} shows that resistance $r$ is determined by the
magnitude of the disorder potential rather than its gradients. Similar to
the case of two wires in series, considered in
Sec.~\ref{Sec-Transport-TwoWires}, this feature can be traced back to the
fact that in the inhomogeneous wire electrons never reach full
equilibration no matter how long the wire is. It is also interesting to
note that resistance of the wire in the regime of strong equilibration
[Eq.~\eqref{R-3}] can be understood from purely hydrodynamic
considerations.~\cite{AS}

\begin{figure}
  \includegraphics[width=8cm]{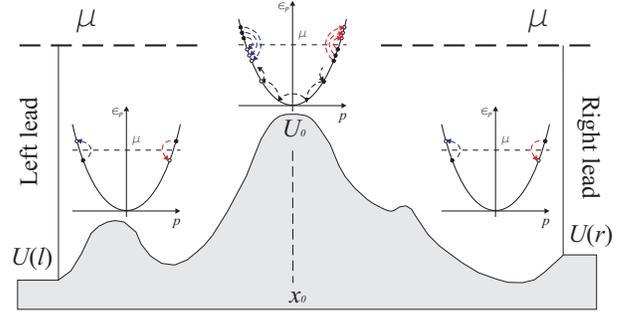}
  \caption{[Color online] Enhanced equilibration due to three-particle collisions in the
   wire segment of length $\ell_T$ where the inhomogeneity potential is
   maximal. A larger value of $e^{-\mu(x)/T}$ near $x\sim x_0$ favors
   stronger electron backscattering in accordance with Eq.~\eqref{NR-kinetic}.
   In contrast, momentum-non-conserving two-particle collisions occur throughout the wire. }\label{Fig-Barrier}
\end{figure}

\subsubsection{Strong disorder}

We now relax the assumption of small variations in the amplitude of
$U(x)$. Consider the case when generally smooth profile of the
inhomogeneity potential has a well-defined maximum $U_0\gg T$ at some
point $x_0$ inside the wire, see Fig.~\ref{Fig-Barrier}. We assume that
such strong fluctuation of $U(x)$ happens in only one place along
the wire. This situation is likely to occur in practice due to the
presence of charged impurities in the substrate and/or uneven screening of
the nearby gates.

Since $e^{-\mu(x)/T}$ is largest near $x_0$ and
$d\dot{N}^R/dx\propto e^{-\mu(x)/T}$, see Eq.~\eqref{NR-kinetic}, it
is natural to expect that three-particle equilibration processes are
significantly enhanced in the part of the wire where $U(x)$ reaches
its maximum. Thus this region of the wire gives dominant
contribution to resistance at lowest temperatures when $\gamma^2\gg
r_1\overline{\mu}^2\gg r_1\overline{\delta\mu^2}$. Assuming
smoothness of $\ell_{1}(x)$ at $x\sim x_0$ as compared to the sharp
$e^{-\mu(x)/T}$ we can compute the resistance $r=r_1$by applying the
saddle point approximation to Eq.~\eqref{r1-average},
\begin{equation}\label{r1-saddle-point}
r=\frac{\ell_{T}}{\ell_{1}}e^{-\mu_0/T}\,,\quad
\ell_{T}=\sqrt{\frac{2\pi T}{|U''(x_0)|}}.
\end{equation}
Here $\mu_0=\mu-U(x_0)$, $\ell_1=\ell_{1}(x_0)$ and $\ell_{T}$ is
the thermal length associated with the curvature of the potential
$U(x)$ near $x_0$.

In contrast to the equilibration processes, dominated by scattering
near $x_0$, momentum-non-conserving two-particle collisions occur
throughout the wire. Resistance of the wire is controlled by the
latter at intermediate temperatures when
$r_1\overline{\mu}^2\gg\gamma^2\gg r_1\overline{\delta\mu^2}$. For
this regime
\begin{equation}\label{r2-saddle-point}
r=\frac{TL}{16\mu^2_0}\left\langle\frac{\Upsilon(x)\mu(x)}{n(x)}\right\rangle\,,
\end{equation}
which is analogous to Eq.~\eqref{R-2}. The only difference is that
due to strong variations of $U(x)$ spatial averaging in
Eq.~\eqref{r2-saddle-point} involves not only $\Upsilon(x)$.

At higher temperatures, when $ r_1\overline{\mu}^2\gg
r_1\overline{\delta\mu^2}\gg\gamma^2$, resistance is again dominated by
the scattering processes near the top of the inhomogeneity potential. In
this regime the first term in Eq.~\eqref{R-crossover-2}, determined by the
amplitude fluctuations of $U(x)$ rather than its gradient, gives the
leading contribution
\begin{equation}\label{r3-saddle-point}
r=\frac{\pi^2}{24}\frac{T^4}{\mu^4_0}\frac{\ell_T}{\ell_{\mathrm{eq}}(x_0)}\,.
\end{equation}
Here we used $\overline{\delta\mu^2}/\overline{\mu}^2=T^2/2\mu^2_0$
found within saddle-point approximation from Eqs.~\eqref{mu-average}
and \eqref{mu2-average} to leading order in $T/\mu_0\ll1$. Equation
\eqref{r3-saddle-point} is analogous to Eq.~\eqref{R-3}, with the
thermal length $\ell_T$ effectively playing the role of the system
size.

%------------------------------------------------------------------------------------------------------------
\section{Summary}\label{Sec-Discussions}

In this paper we studied the transport properties of weakly
interacting one-dimensional electrons in the presence of
inhomogeneities. In this system equilibration is strongly restricted
by the phase space available for electron scattering and
conservation laws. The resulting equilibration length
$\ell_{\mathrm{eq}}\propto e^{\mu/T}$ is exponentially large at low
temperatures and the partially equilibrated state is more likely to
be realized than the fully equilibrated one. Furthermore,
inhomogeneities present in the wire themselves resist equilibration
of electrons due to momentum-non-conserving two-particle collisions.

Our main results are expressions \eqref{R}--\eqref{S} for the
resistance, Peltier coefficient, thermal conductance and
thermopower. We find that the combined effect of interactions and
inhomogeneities can dramatically increase the resistance of the
wire. For the long enough wire the induced correction could be much
greater than the contact resistance of noninteracting electrons
$h/2e^2$. This is in contrast to the uniform case where
interaction-induced correction to resistance saturates for
$L\gg\ell_{\mathrm{eq}}$ and remains small as $(T/\mu)^2$
compared to the resistance $h/2e^2$ of a non-interacting wire.

The combined effect of interactions and inhomogeneities is different
for thermoelectric coefficients. On the one hand, when temperature
increases, Peltier coefficient and thermopower grow from
exponentially small values, Eq.~\eqref{Pi-S-0}, to
$\Pi=TS\simeq(\pi^2 T^2/6e)(\overline{\mu}/\overline{\mu^2})$. On
the other, this enhancement is not as dramatic as in the case of
resistance. Indeed, the difference between the saturated values of
Peltier coefficient and thermopower in inhomogeneous wire as
compared to that in the uniform wires is only in the appearance of
the renormalized factor $\overline{\mu}/\overline{\mu^2}$ instead of
the inverse chemical potential $1/\mu$, see Eq.~\eqref{Pi-S-eq}.
Conversely, the thermal conductance of the wire decreases due to the
equilibration from its noninteracting value, Eq.~\eqref{K-0}, to
zero at $L\gg\ell_{\mathrm{eq}}$.

The lack of complete electronic equilibration in the inhomogeneous
quantum wires, which is another central observation of our work,
warrants additional discussion. The notion of equilibration appears
naturally since initially right- and left-moving electrons entering
the wire from the left and right lead, respectively, are at
different equilibria with respect to each other due to the applied
bias or temperature difference. In the case of weak interactions
three-particle collisions constitute the leading-order relaxation
process. Although the corresponding relaxation rate is slow (or
equivalently relaxation length is large) due to the required
scattering through the bottom of the band, complete equilibration
between the right and left movers is nevertheless possible in a
\textit{homogeneous} wire. Once the length of the wire becomes large
such that exponential suppression of the equilibration effects is
compensated by a large system size, $L\gg\ell_{\mathrm{eq}}\sim
e^{\mu/T}$, the relaxation of the electron system becomes
significant. Right and left movers eventually equilibrate to the
single distribution, Eq.~\eqref{f-eq}. When viewed in a reference
frame moving together with electrons this distribution is simply the
equilibrium Fermi function. Thus, in the stationary frame this is
distribution with a boost.

In the presence of spatial {\em inhomogeneities} full equilibration
is impeded. This is most transparent when momentum-nonconserving
two-body collisions are present, which unlike three-particle
processes favor electron distribution function without boost,
Eq.~\eqref{f-0}. As a result, the electron system is frustrated due
to competition between two scattering processes and an intermediate
distribution, Eq.~\eqref{f-part-eq}, is established.

Interestingly, lack of full equilibration is a more general
characteristic of inhomogeneous quantum wires, which is a
consequence of conservation laws. More precisely, in inhomogeneous
wires it is generally not possible to reconcile full equilibration
with conservation of energy. Technically speaking, this observation
comes from the fact that in the fully equilibrated state all
currents are proportional to the drift velocity. In the linear
response regime one therefore has only a single parameter $v_d$ to
simultaneously satisfy uniformity of particle and heat currents
along the wire, imposed by conservation laws. In inhomogeneous wires
this is generally not possible since for $\Delta\mu=0$ the ratio
$j_Q/I=\pi^2T^2/6e\mu(x)$ is not constant along the wire. Thus the
electron liquid must remain in the state of partial equilibration
with $\Delta\mu\neq0$. We have illustrated this point explicitly by
considering the simplest example of inhomogeneity, namely, a
junction of two uniform wires with mismatched densities. In the
general case of a wire with long-range disorder the consequence of
the partially equilibrated state is that interaction-induced
correction to resistance of the wire is determined by the amplitude
of the variations in inhomogeneity potential, rather than its
gradients. This correction may be large compared to the resistance
$h/2e^2$ of non-interacting wires.

%-------------------------------------------------------------------------------------------------------------

\subsection*{Acknowledgement}

We are grateful to B.~L.~Altshuler, A.~V.~Andreev, A.~P.~Dmitriev,
Y.~M.~Galperin, I.~V.~Gornyi, D.~G.~Polyakov and B.~Shklovskii for
helpful discussions. This work at ANL was supported by the U.S.
Department of Energy, Office of Science, under Contract No.
DE-AC02-06CH11357.

%-------------------------------------------------------------------------------------------------------------
\appendix\label{Sec-Appendix}

%---------------------------------------------------------------------------------------------------------
\section{Details on the calculation of
$\dot{Q}^{R}_{p}$}\label{Sec-Appendix-QR}

For the inhomogeneous wire we describe the electron-electron
interaction responsible for two-body scattering by its general
translationally non-invariant form,
\begin{equation}
\mathcal{V}(x,x')\Rightarrow
\mathcal{V}\left(x-x',\frac{x+x'}{2}\right)\,.
\end{equation}
As a function of its first argument $\mathcal{V}$ is assumed to be
Coulombic in nature and thus short-ranged, with variations on the
scale of a certain screening length due to the nearby gates. The
inhomogeneity is captured by the second argument with a
corresponding variations in $\mathcal{V}$ on the length scale $b$,
large compared to both, the Fermi wavelength $\lambda_F$ and the
range of screening. Our starting point for the energy-transfer rate
from the right movers for the segment of the wire $\hbar
v_F/T\ll\Delta x\ll b$ is the following golden-rule expression,
\begin{eqnarray}\label{QRp-GoldenRule}
&&\Delta\dot{Q}^R_p(x)=-\frac{2\pi}{\hbar}\int\frac{d\epsilon_p
d\epsilon_{p'}d\epsilon_k
d\epsilon_{k'}}{(2\pi)^4}|\mathcal{V}(\epsilon_p,\epsilon_k;\epsilon_{p'},\epsilon_{k'})|^2
\nonumber\\
&&\delta(\epsilon_p+\epsilon_k-\epsilon_{p'}-\epsilon_{k'})
(\epsilon_{p}-\epsilon_{p'}+\epsilon_{k'}-\epsilon_k)
\nonumber\\
&&\left[f^{R}(\epsilon_p)(1-f^{R}(\epsilon_{p'}))f^{L}(\epsilon_k)(1-f^{L}(\epsilon_{k'}))-\right.
\nonumber\\
&&\left.f^{R}(\epsilon_{p'})(1-f^{R}(\epsilon_{p}))f^{L}(\epsilon_{k'})(1-f^{L}(\epsilon_{k}))\right]\,,
\end{eqnarray}
where the matrix element
\begin{equation}
|\mathcal{V}|^2=|\mathcal{V}_{\parallel}|^2+
|\mathcal{V}_{1\perp}|^2+ |\mathcal{V}_{2\perp}|^2\,,
\end{equation}
includes three possible scattering processes of spin-full electrons
from the initial state $(\epsilon_p,\epsilon_k)$ into the final
state $(\epsilon_{p'},\epsilon_{k'})$. These matrix elements will be
calculated on the basis of semiclassical wave functions,
\begin{equation}\label{psi}
\psi_{\epsilon,\pm}(x)=\frac{1}{\sqrt{\hbar
v_{F}(x)}}\exp\left[\pm\frac{i}{\hbar}\int^{x}_{0}dx'\sqrt{2m[\epsilon-U(x)]}\right]\,,
\end{equation}
which are eigenstates of the free Hamiltonian
$[-\hbar^2\partial^{2}_{x}/2m+U(x)]\psi_{\epsilon}(x)=\epsilon\psi_{\epsilon}(x)$
normalized according to $\int
dx\,\psi_{\epsilon,\pm}(x)\psi^{*}_{\epsilon',\pm}(x)=2\pi\delta(\epsilon-\epsilon')$
and $v_{\epsilon}(x)=p_{\epsilon}(x)/m$. The subscripts $\pm$ refer
to the right/left branches and we also ignored the backscattering
wave, since it only leads to an exponentially small contribution for
$k_Fb\gg1$. Focusing on the states close to the Fermi energy, we can
simplify the expression for the eigenstates, Eq.~\eqref{psi}, of the
free Hamiltonian into
\begin{equation}\label{psi}
\psi_{\epsilon,\pm}(x)\approx\psi_{\mu,\pm}(x)\exp\left[\pm
i(\epsilon-\mu)\int^{x}_{0}\frac{dx'}{\hbar v_{F}(x')}\right]\,,
\end{equation}
where $\psi_{\mu,\pm}(x)$ is obtained from Eq.~\eqref{psi} by
setting $\epsilon=\mu$. This allows us to find, for example, the
matrix element $\mathcal{V}_{\parallel}$ to first order in the
interaction
\begin{eqnarray}
\mathcal{V}_{\parallel}(\epsilon_p,\epsilon_k;\epsilon_{p'},\epsilon_{k'})=&&\hskip-0.45cm
\int^{x+\Delta x}_{x}\!\!dX
\frac{\exp\left(i\int^{X}_{0}dx'\frac{\epsilon_{p'}-\epsilon_p+\epsilon_k-\epsilon_{k'}}{\hbar
v_{F}(x')}\right)}{[\hbar v_{F}(X)]^2}
\nonumber\\
&&\hskip-1.25cm\times\int^{\Delta x}_{-\Delta x}dy\,
\mathcal{V}(y,X)\left(1-e^{-2ik_{F}(X)y}\right)\,.
\end{eqnarray}
Here we introduced center of mass $X=(x+x')/2$ and relative $y=x-x'$
coordinates, and $k_{F}(X)=p_{F}(X)/\hbar$.

Using Eq.~\eqref{f-part-eq} and expanding the occupation factors
$f^{R/L}(\epsilon)$ in Eq.~\eqref{QRp-GoldenRule} to linear order in
$u(x)$, and splitting the energy-conserving delta function into two
as $\int d\omega
\delta(\epsilon_p-\epsilon_{p'}-\omega)\delta(\epsilon_{k}-\epsilon_{k'}+\omega)$,
we can complete $\epsilon_{p'}$ and $\epsilon_{k'}$ integrations and
find
\begin{eqnarray}
\Delta\dot{Q}^{R}_{p\parallel}(x)=-\frac{2\pi T u(x)}{\hbar
v_{F}(x)}\int\frac{d\omega
d\epsilon_pd\epsilon_k}{(2\pi)^4}|\mathcal{V}_{\parallel}(\omega)|^2
\nonumber\\
\frac{4\omega^2}{T^2}
f(\epsilon_p)(1-f(\epsilon_p-\omega))f(\epsilon_k)(1-f(\epsilon_k+\omega))\,,
\end{eqnarray}
where $f(\epsilon)$ is now the equilibrium Fermi function. The
corresponding matrix element in these notations reads
\begin{eqnarray}
\mathcal{V}_{\parallel}(\omega)=\int^{x+\Delta
x}_{x}dX\,\frac{\mathcal{V}_0(X)-\mathcal{V}_{2k_F}(X)}{[\hbar
v_{F}(X)]^2}\nonumber
\\
\times\exp\left(-2i\omega\int^{X}_{0}\frac{dx'}{\hbar
v_{F}(x')}\right)\,.
\end{eqnarray}
The shortened forms $\mathcal{V}_0$ and $\mathcal{V}_{2k_F}$
correspond to the zero momentum and $2k_F$ Fourier components of the
potential $\mathcal{V}(y,X)$ with respect to its first variable $y$
defined as $\mathcal{V}_{0}(X)=\int dy\,\mathcal{V}(y,X)$ and
$\mathcal{V}_{2k_F}(X)=\int dy\,\mathcal{V}(y,X)e^{-2ik_{F}(X)y}$.
At this stage $\epsilon_p$ and $\epsilon_k$ integrations can be
completed by noticing that
\begin{eqnarray}
f(\epsilon)(1-f(\epsilon\pm\omega))=\frac{f(\epsilon)-f(\epsilon\pm\omega)}{1-e^{\mp\omega/T}}\,,
\nonumber\\
\int^{+\infty}_{-\infty} d\epsilon\,
[f(\epsilon)-f(\epsilon\pm\omega)]=\pm\omega\,.
\end{eqnarray}
As a result, we obtain following expression for the energy-transfer
rate:
\begin{eqnarray}
&&\hskip-1cm\Delta\dot{Q}^R_{p\parallel}(x)=-\frac{Tu(x)}{8\pi\hbar
v_{F}(x)}\iint^{x+\Delta x}_{x}dX_1dX_2
\nonumber\\
&&\hskip-1cm\left[\frac{\mathcal{V}_{0}(X_1)-\mathcal{V}_{2k_F}(X_1)}{\pi\hbar
v_{F}(X_1)}\right]\left[\frac{\mathcal{V}_{0}(X_2)-\mathcal{V}_{2k_F}(X_2)}{\pi\hbar
v_{F}(X_2)}\right]
\nonumber\\
&&\hskip-1cm\int^{+\infty}_{-\infty}d\omega\left(\frac{\omega^2/T}{\sinh\frac{\omega}{2T}}\right)^2
\frac{\exp\left(-2i\omega\int^{X_2}_{X_1}\frac{dx'}{\hbar
v_F(x')}\right)}{\hbar v_{F}(X_1)\hbar v_{F}(X_2)} \,.
\end{eqnarray}
Being interested in the temperature range $T\gg\hbar v_F/b$, where
the exponential is rapidly oscillating, we write
\begin{eqnarray}
\frac{4\omega^2}{\hbar v_{F}(X_1)\hbar v_{F}(X_2)}
\exp\left(-2i\omega\int^{X_2}_{X_1}\frac{dx'}{\hbar
v_F(x')}\right)\nonumber
\\
=\frac{\partial^2}{\partial X_1\partial
X_2}\exp\left(-2i\omega\int^{X_2}_{X_1}\frac{dx'}{\hbar
v_F(x')}\right)\,,
\end{eqnarray}
integrate by parts over $X_1$ and $X_2$ and complete the remaining
energy integral, which gives $4\pi\hbar v_{F}(x)\delta(X_1-X_2)$.
Due to the delta function, one spatial integration is thus removed
and we find as the final result, Eq.~\eqref{QRp}--\eqref{l-gamma},
where all three scattering channels were included.

%---------------------------------------------------------------------------------------------------------
\section{Series-resistance of two uniform wires from the conservation
laws}\label{Sec-Appendix-R}

The resistance of a junction between two uniform wires with
different densities can be found simply by combining conservation
laws for the currents with the microscopic equations for $\dot{N}^R$
and $\dot{Q}^R$. Our starting point is Eq.~\eqref{Int-Landauer}
which is naturally generalized to the case of two wires connected in
series
\begin{equation}\label{Int-Landauer-two-wires}
\frac{2e^2V}{h}=I-e\dot{N}^{R}_{1}-e\dot{N}^{R}_{2}\,,
\end{equation}
where $\dot{N}^R_{1,2}$ correspond to the change in the number of
right-movers within each segment of the wire. Similarly,
generalization of Eq.~\eqref{Heat-balance} for the heat balance
reads
\begin{equation}\label{Heat-balance-two-wires}
\frac{2\pi^2T\Delta T}{3h}=j_Q-\dot{Q}^R_1-\dot{Q}^R_2\,.
\end{equation}
Equation \eqref{Int-Landauer-two-wires} defines the resistance,
Eq.~\eqref{R}, where interactions-induced correction is given by
\begin{equation}\label{r-two-wires-def}
r=-\frac{e\dot{N}^{R}_{1}}{I}-\frac{e\dot{N}^{R}_{2}}{I}\,.
\end{equation}
We use now Eq.~\eqref{NR-kinetic} in the limit of uniform wires to
express rates $\dot{N}^R_{1,2}$ as
\begin{equation}\label{NR-two-wires}
e\dot{N}^R_{i}=-I(1-\alpha_{i})\frac{L_{i}}{\lambda_{i}}\,,\quad
\lambda_{i}=\ell^{(i)}_{1}e^{\mu_{i}/T}\,,\quad i=1,2\,,
\end{equation}
by introducing the new quantity $\alpha_i=u_i/v_d$, which has
meaning of the degree of equilibration. Indeed, $\alpha=0$
corresponds to the limit of no equilibration, such that the current
is determined solely by $\Delta\mu$, while $\alpha=1$ corresponds to
full equilibration where $I=env_d$. This notation makes it possible
to express $\Delta\mu_i$ in Eq.~\eqref{I} in terms of $\alpha_i$ as
$I=2e\Delta\mu_i/h+\alpha_iI$, which was used in
Eq.~\eqref{NR-two-wires}.

Conservation of the heat current, Eq.~\eqref{jQ},
\begin{equation}\label{jQ-two-wires}
ej_Q=\frac{\pi^2}{6}\frac{T^2}{\mu_1}\alpha_1I=\frac{\pi^2}{6}\frac{T^2}{\mu_2}\alpha_2I
\end{equation}
imposes a constraint $\alpha_1/\mu_1=\alpha_2/\mu_2\equiv\chi$,
which must be constant. With the help of Eq.~\eqref{NR-two-wires}
the expression for $r$ in Eq.~\eqref{r-two-wires-def} can be
rewritten as
\begin{equation}\label{R-two-wires-2}
r=\frac{L_1}{\lambda_1}+\frac{L_2}{\lambda_2}-
\chi\left(\mu_1\frac{L_1}{\lambda_1}+\mu_2\frac{L_2}{\lambda_2}\right)\,.
\end{equation}
where the unknown quantity $\chi$ is yet to be determined. The way
to find it is from the energy conservation. Recall, that $\dot{Q}^R$
and $\dot{N}^R$ are related to each other by Eq.~\eqref{QRb} since
they are caused by the same scattering mechanism. For the uniform
wire Eq.~\eqref{QRb} reads $\dot{Q}^R_{i}=-2\mu_i\dot{N}^R_i$. As a
result, for Eq.~\eqref{Heat-balance-two-wires} in the case of no
temperature bias $(\Delta T=0)$ we have
\begin{equation}
j_Q+2\mu_1\dot{N}^R_1+2\mu_2\dot{N}^R_2=0\,.
\end{equation}
Inserting here Eqs.~\eqref{NR-two-wires} and \eqref{jQ-two-wires}
one finds
\begin{equation}
\frac{\pi^2}{6}T^2\chi-2\mu_1(1-\chi\mu_1)\frac{L_1}{\lambda_1}-
2\mu_2(1-\chi\mu_2)\frac{L_2}{\lambda_2}=0\,,
\end{equation}
which allows to find $\chi$ explicitly,
\begin{equation}
\chi=\frac{2\mu_1\frac{L_1}{\lambda_1}+2\mu_2\frac{L_2}{\lambda_2}}
{\frac{\pi^2T^2}{6}+2\mu^2_1\frac{L_1}{\lambda_1}+2\mu^2_2\frac{L_2}{\lambda_2}}\,.
\end{equation}
Finally, using this $\chi$ in Eq.~\eqref{R-two-wires-2} and
reexpressing $\lambda$ through the equilibration length as
$\lambda_i=(12\mu^2_i/\pi^2T^2)\ell^{(i)}_{\mathrm{eq}}$ one
recovers Eq.~\eqref{R-two-wires}. We thus conclude that
Eq.~\eqref{R-two-wires} is just a consequence of the conservation
laws.

%---------------------------------------------------------------------------------------------------------
%---------------------------------------------------------------------------------------------------------

\end{document}